\documentclass[preprint,aps,showpacs,nofootinbib,preprintnumbers,amsmath,amssymb]{revtex4}
\usepackage{amssymb}
\usepackage{}
\usepackage{epsfig}
\usepackage{graphicx}
\usepackage{subfigure}
\usepackage{dcolumn}
\usepackage{bm}
\usepackage[usenames ,dvipsnames]{xcolor}
\usepackage{slashed}

\begin{document}

\title{Constraints on the septet-doublet mixing models from oblique parameters}

\author{Chao-Qiang~Geng$^{1,2,3}$\footnote{geng@phys.nthu.edu.tw},
Lu-Hsing Tsai$^{1}$\footnote{lhtsai@phys.nthu.edu.tw}, and Yao Yu$^{3}$\footnote{yuyao@cqupt.edu.cn}}
  \affiliation{$^{1}$Chongqing University of Posts \& Telecommunications, Chongqing, 400065, China\\
  $^{2}$Department of Physics, National Tsing Hua University, Hsinchu, 300, Taiwan\\
$^{3}$Physics Division, National Center for Theoretical Sciences, Hsinchu, 300, Taiwan}

\date{\today}
\begin{abstract}
The limitations of the doublet-septet mixing models by the deviations of electroweak oblique parameters $\Delta S$ and $\Delta T$ are studied. 
In the minimal model, the mixture of  the septet $\eta$ and the scalar doublet in the standard model (SM) is driven by a 
non-Hermitian dimension-7 operator. For a smaller bare mass of the septet, $\Delta S$ gives a stringent constraint on  the mixing angle 
$\sin\beta$ between the CP-odd neutral parts of the SM Higgs doublet and $\eta$. In general,  increasing the mass of the scalar septet $M_\eta$ will 
enhance the deviation of $T$ from the SM, whereas it decreases the magnitude of $\Delta S$ for a larger bare mass within 
the range $M_\eta\lesssim 400\,{\rm GeV}$. We also examine two extended models from the ordinary doublet-septet 
mixture pattern. One of them is based on a inert doublet-septet mixing pattern, in which there is no vacuum expectation 
value for the neutral component of $\eta$, and a stable dark matter could naturally exist. For a benchmark point 
with  this inner doublet mass of $M_\chi=250{\rm}$ and $M_\eta=400\,{\rm GeV}$ in this model, the mixing coefficient is found to be less than $1.8$. 
The other extension is constructed by imposing a doubly charged scalar mixed with the doubly charged component 
of the septet. Apart from the contribution by the septet-doublet admixture, $\Delta S$ is suppressed by a factor of 
$s_W^2$ and $\Delta T$ has a significant constraint due to the vanishing vacuum polarization of $Z$  at the momentum transfer $p^2=0$.
\end{abstract}

%
\maketitle
\section{Introduction}

The discovery of the $125\,{\rm GeV}$ Higgs boson makes the standard model (SM) complete~\cite{atlas:2012gk, cms:2012gu}. 
However, there is still a possibility that some fraction of this light boson comes from other unknown scalar particles,
 carrying quantum numbers $(I, Y)$ under the gauge symmetry of ${\rm SU}(2)_L\times {\rm U}(1)_Y$, 
 for which the electric charge is related by $Q=I_3+Y/2$, where $I_3$ is the third component the of weak isospin $I=(n-1)/2$,  with $n$ being the dimension of the representation for ${\rm SU}(2)_L$. 
 One example is to impose one or more scalar doublets with the quantum number $(2,1)$~\cite{Lee:1973iz, Branco:2011iw} 
 or singlet $(1,0)$~\cite{Davoudiasl:2004be, O'Connell:2006wi} to couple with the ordinary SM scalar doublet. 
 It is interesting to note that the vacuum expectation value (VEV) of the singlet or doublet does not change the $\rho$ parameter 
 from unity at tree level~\cite{Ross:1975fq, Veltman:1977kh}, so that these types of the models can only be constrained by the electroweak oblique parameters of $S$, $T$, and $U$~\cite{Peskin:1990zt, Peskin:1991sw}, in which $S=-0.03\pm0.10$, $T=0.01\pm0.12$, and $U=0.05\pm0.10$
have been given by  the recently global fitting~\cite{Agashe:2014kda}. In particular, it has been pointed out that the mixing angles in singlet-doublet mixing~\cite{Barger:2007im, Lopez-Val:2014jva}, two Higgs doublet~\cite{Toussaint:1978zm, Bertolini:1985ia, Kanemura:2011sj, LopezVal:2012zb}, and multi-doublet~\cite{Grimus:2007if, Dawson:2009yx} models are bounded by $S$ and $T$.

Besides the SU(2) singlet or doublet, there are a series of specific higher multiplets which can also retain $\rho=1$, 
In particular, the SU(2) septet $\eta$ with $Y=4$ is the smallest choice of the multiplet to have the feature~\cite{Hally:2012pu}. 
However, the septet with a nonzero VEV via the renormalizable operators is not allowed. 
Instead, some higher dimensional operators involving the septet and doublet are required to offer the mixings between them~\cite{Hisano:2013sn}. Although the VEV of the  septet $v_\eta$ is not limited by the $\rho$ parameter, electroweak oblique parameters could further constrain it to be $v_\eta\lesssim 20\,{\rm GeV}$~\cite{Kanemura:2013mc}. Furthermore, the Higgs-gauge and Higgs-fermion couplings observed from the LHC also restrict the structure of the septet~\cite{Killick:2013mya, Kanemura:2014bqa} with  $v_\eta\lesssim 6\,{\rm GeV}$~\cite{Alvarado:2014jva}.

There are also many applications for the scenarios that the extra multiplet cannot preserve the unity for $\rho$. The value of $\rho=1.0000\pm0.0009$~\cite{Agashe:2014kda} from the global fitting constrains the VEV of the multiplet up to order of several GeV, which means that the multiplet should be inert from the SM Higgs doublet, with only a tiny mixture allowed. The Higgs triplet model (Type-II seesaw)~\cite{typeIIseesaw1, typeIIseesaw2, typeIIseesaw3, typeIIseesaw4, typeIIseesaw5, typeIIseesaw6, typeIIseesaw7} is one of the typical example in which the VEV of the triplet is limited to be less than $\mathcal{O}(1)$ GeV~\cite{Gunion:1989in, Chun:2003ej, Melfo:2011nx, Chun:2012jw, Chun:2013vca}. The constraint on the VEV of the quintuplet $(5,2)$ was also studied~\cite{Earl:2013jsa}. For the extreme case that the VEV of the multiplet is forbidden by some discrete or continue symmetry, the oblique parameters can help to constrain the mixings of the multiplet with other particles. This type of the models can also contain dark matter if there exists a lightest component carrying a nonzero charge for the symmetry~\cite{Cirelli:2005uq, Cirelli:2009uv, Hambye:2009pw, Cai:2012kt, Earl:2013fpa}. For more complicated situations, it is worth to explore the possibility that one or more scalar doublets and singlets mix
 with the septet.

This paper is organized as follows. In Sec.~II, we review the septet-doublet mixing model and constrain the model from the oblique parameters. In Sec.~III, we study the extended septet models. We give our conclusions in Sec.~IV.

\section{The doubly-septet mixing model}
It is known that the non-zero VEVs of new SU(2)$\times$U(1) multiplets with $(n,Y)$ could contribute to $\rho$ 
with the general form~\cite{Lee:1972zzc}
\begin{eqnarray}
\label{eq1}
\rho\equiv {m_W^2\over m_Z^2 c_W^2}={\sum_i[I_{(i)}(I_{(i)}+1)-{1\over4}Y_{(i)}^2]v_{(i)}^2\over \sum_i{1\over2}Y_{(i)}^2v_{(i)}^2}\;,
\end{eqnarray}
where 
the subscript $(i)$ indicates for the $i$-th multiplet.
 From Eq.~(\ref{eq1}), it is easy to see that both singlet and doublet scalars as well as the septet $\eta :(7,4)$ will not change 
 the $\rho$ parameter from unity. 
 We would like to examine some models with the septet  to keep $\rho=1$ but constrained by the oblique parameters.
 
 We start with a simple model in which $\eta$ acquires a non-zero VEV and has a mixing with the SM doublet.
This model including two scalar components: the scalar doublet $\Phi=(\Phi^+\;,\;\Phi^0)^T$ in the SM, and the septet $\eta$, which has the irreducible form consisting of seven independent complex components with the electric charges from $Q=+5$ to $-1$, given by
\begin{eqnarray}
\eta=(\eta^{(+5)}\;,\;\eta^{(+4)}\;,\;\eta^{(+3)}\;,\;\eta^{++}\;,\;\eta^{+}\;,\;\eta^{0}\;,\;\eta^{-})\;.
\end{eqnarray}
The relevant terms of the scalar potential in this model are then given by
\begin{eqnarray}
\label{eq3}
-L=-\mu^2(\Phi^*\Phi)+\lambda (\Phi^*\Phi)^2+M_\eta^2(\eta^*\eta)+\Big[{1\over \Lambda^3}\eta \Phi(\Phi^*)^5+{\rm H.c.}\Big]\;,
\end{eqnarray}
where the terms proportional to $\mu$ and $\lambda$ are the couplings in the SM, $M_\eta$ is the bare mass of $\eta$, and
and $\Lambda$ is an effective energy scale. Note that the dimension-7 effective operator $(1/ \Lambda^3)\eta \Phi(\Phi^*)^5$ in Eq.~(\ref{eq3}) is the simplest form of a higher dimensional operator containing a mixture of $\eta$ and $\Phi$ which can generate nonzero VEV for $\eta$ at the low energy scale~\cite{Hisano:2013sn}.  Here, we also assume that there is no other lower dimensional effective operator. 
 Other renormalizable gauge invariant terms such as $(\Phi^*\Phi\eta^*\eta)$ and ($\eta\eta\eta^*\eta^*$) are not relevant in this paper and will be ignored in this study.

After the spontaneous symmetry broken due to the negative quadratic term of the scalar doublet, it leads to the VEV with $\Phi^0=(v_\Phi+R_\Phi+iI_\Phi)/\sqrt{2}$, and the the neutral part of the septet also acquires the VEV with $\eta^0=(v_\eta+R_\eta+iI_\eta)/\sqrt{2}$ via the dimension-7 operator,  with $v_\Phi$($v_\eta$), $R_\Phi$($R_\eta$), and $I_\Phi$($I_\eta$) being the VEV, CP-even component, and CP-odd component of $\Phi^0$($\eta^0$), respectively. Then, we can take all mixing states of the doublet and septet into account, including neutral, singly charge, and doubly charged states. 
We discuss the mass spectrum of the septet by following the formulae in Ref.~\cite{Hisano:2013sn}. 
In general, the weak and mass eigenstates of the scalars can be expressed by 
\begin{eqnarray}
\left(
\begin{array}{c}
R_\Phi\\
R_\eta\\
\end{array}\right)
=
\left(
\begin{array}{cc}
\cos\alpha&-\sin\alpha\\
\sin\alpha&\cos\alpha\\
\end{array}\right)
\left(
\begin{array}{c}
h\\
H\\
\end{array}\right)\;,\;
\left(
\begin{array}{c}
I_\Phi\\
I_\eta\\
\end{array}\right)
=
\left(
\begin{array}{cc}
\cos\beta&-\sin\beta\\
\sin\beta&\cos\beta\\
\end{array}\right)
\left(
\begin{array}{c}
G^0\\
A^0\\
\end{array}\right)\;,\;\label{Eq_Neutralmixing}
\end{eqnarray}

\begin{eqnarray}
\left(
\begin{array}{c}
\Phi^+\\
\eta^+\\
(\eta^-)^*\\
\end{array}\right)
=
U
\left(
\begin{array}{c}
G^+\\
\bar\eta^+\\
S^+\\
\end{array}\right)\;,\;
\end{eqnarray}
where  $h$, $H$, $G^0$ and $A^0$ are the neutral mass eigenstates; $G^+$, $\bar\eta^+$ and $S^+$ are the singly charged mass eigenstates; $\alpha$ and $\beta$ are the mixing angles\footnote{ After taking $\beta\rightarrow (\pi/2)-\beta$ and $\alpha\rightarrow -\alpha$, the notations become those in Ref.~\cite{Hisano:2013sn}.} with $\tan\beta=(4v_\eta)/v_\Phi$,
and  the $3\times3$ matrix $U$ is given by
\begin{eqnarray}
U=\left(\begin{array}{ccc}
c_\beta&0&s_\beta\\
{\sqrt{10}\over4}s_\beta&\sqrt{6}\over4 &-{\sqrt{10}\over4}c_\beta \\
-{\sqrt{6}\over4}s_\beta&\sqrt{10}\over4 &{\sqrt{6}\over4}c_\beta \\
\end{array}\right)\;.
\end{eqnarray}
The mass spectra of the scalars are obtained as
\begin{eqnarray}
M_h^2&=&(1-{3\over2}{t_\beta\over t_\alpha})M_\eta^2\;,\;
M_H^2=(1+{3\over2}t_\alpha t_\beta)M_\eta^2\;,\;M_{A^0}={M_\eta\over c_\beta}\;,\nonumber\\
M_{\bar\eta^\pm}&=&M_\eta\;,\;M_{S^\pm}={M_\eta\over c_\beta}\;,\;\nonumber\\
M_{\eta^{+5}}&=& M_{\eta^{+4}}= M_{\eta^{+3}}=M_{\eta^{+2}}=M_\eta\;,\;\label{Eq_MassEigenvalues}
\end{eqnarray}
where $s_\theta\equiv\sin \theta$, $c_\theta\equiv\cos \theta$, and $t_\theta\equiv\tan \theta$. The related details for the mass matrices are given in Appendix~\ref{Sec_AppendixMixing}. Without the mixing term $\eta \Phi(\Phi^*)^5$, $\beta$ becomes zero and there is no mass splitting among the septet. Note that $\alpha$ can be determined once we take $m_h=125.7\,{\rm GeV}$~\cite{Agashe:2014kda} and fix $\sin\beta$. Subsequently, $m_H$ can be evaluated too.

\begin{figure}
\includegraphics[width=10cm]{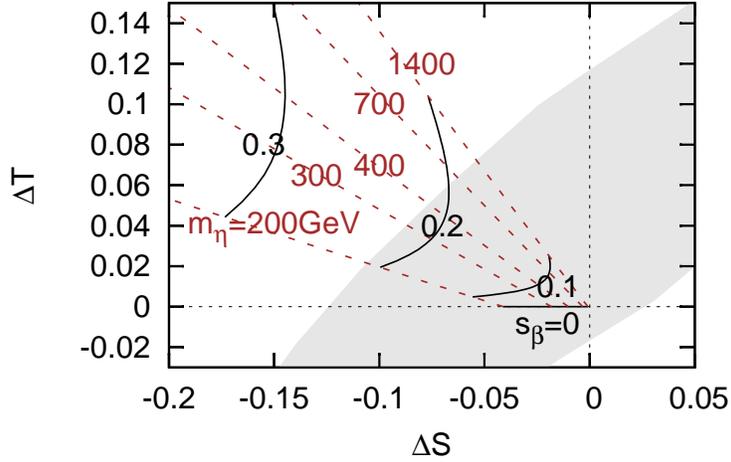}
\caption{Contours for $M_\eta$ and $\tan\beta$ in the $\Delta S-\Delta T$ plane, where the gray region corresponds to the global fitting results at $90\%$ C.L~\cite{Agashe:2014kda}.}
\label{Fig_STplot}
\end{figure}

Recall that the definitions of $S$, $T$, and $U$ parameters are given by~\cite{Maksymyk:1993zm, Lavoura:1993nq}
\begin{eqnarray}
S&=&{16\pi c_W^2 s_W^2\over e^2}\Big[{\Pi_{ZZ}(M_Z^2)-\Pi_{ZZ}(0)\over M_Z^2}
-{c_W^2-s_W^2\over c_Ws_W}{\partial \Pi_{\gamma Z}(p^2)\over\partial (p^2)}\Big|_{p^2=0}
-{\partial \Pi_{\gamma \gamma}(p^2)\over\partial (p^2)}\Big|_{p^2=0}\Big]\;,\\
T&=&{4\pi\over e^2}\Big[{\Pi_{WW}(0)\over M_W^2}-{\Pi_{ZZ}(0)\over M_Z^2}\Big]\;,\\
U&=&{16\pi s_W^2\over e^2}\Big[{\Pi_{WW}(M_W^2)-\Pi_{WW}(0)\over M_W^2}
-c_W^2{\Pi_{ZZ}(M_Z^2)-\Pi_{ZZ}(0)\over M_Z^2}\nonumber\\
&&-2c_Ws_W{\partial \Pi_{\gamma Z}(p^2)\over\partial (p^2)}\Big|_{p^2=0}
-s_W^2{\partial \Pi_{\gamma \gamma}(p^2)\over\partial (p^2)}\Big|_{p^2=0}\Big]\;,
\end{eqnarray}
where $\Pi_{ab}(p^2)$ are the coefficients of $g_{\mu\nu}$ for the vacuum polarizations of gauge bosons $a$ and $b$ ($a,b=W,Z,\gamma$) 
under the momentum transfer $p^2$.
For $\Delta U=0$, the deviation parameters of $\Delta S$ and $\Delta T$ from the data are $\Delta S=0.00\pm0.08$ 
and $\Delta T=0.05\pm0.07$~\cite{Agashe:2014kda}, respectively. 
 We find that in this model $|\Delta U|$ is typically smaller than $|\Delta S|$ and $|\Delta T|$ (see the expression of $\Delta U$ in Appendix \ref{Sec_AppendixST}), with the order of $10^{-3}$ for $M_\eta\gtrsim 200\,{\rm GeV}$ and $\sin\beta\lesssim 0.3$, so that the assumption of 
 $\Delta U=0$ is viable.
We show the formulae of $\Delta S$ and $\Delta T$ in Appendix~\ref{Sec_AppendixST}. Both of them only depend on $M_\eta$ and $\tan\beta$, with the one-to-one correspondence in the interesting area of the $S-T$ plane, in the sense that there is no intersection among the curves 
with a constant $M_\eta$ (or $\tan\beta$) as shown in Fig.~\ref{Fig_STplot}. In general, $\Delta S$ is negative, whereas $\Delta T$ is always positive. Turning off the dimension-7 interaction ($\sin\beta\rightarrow 0$) will go back to an inert septet without a VEV, for which $\Delta T=0$, while $\Delta S$ approaches zero for a large $M_\eta$. Notice that the allowed region by the observation at $90\%$C.L. is positively correlated between $\Delta S$ and $\Delta T$~\cite{Agashe:2014kda}, so that for a larger value of $M_\eta$ the restriction on $\sin\beta$ does not become much more strongly despite its larger contribution to $\Delta T$. As an example, we have $\sin\beta\lesssim 0.21(0.22)$ for $M_\eta=200(300)\,{\rm GeV}$. 
 We note that $M_\eta$ is also limited by the experimental constraint from the pair production of $\eta$, dominated by Drell-Yan processes, which is independent of $v_\eta$ and $\sin\beta$. In Ref.~\cite{Alvarado:2014jva}, $M_\eta$ is found to be larger than $\sim 400\,{\rm GeV}$. Similar results about the oblique parameters in this model were also discussed in Ref.~\cite{Hisano:2013sn}. 
We point out that our results for $\Delta S$ have two features which are different from those in Ref.~\cite{Hisano:2013sn} (see Eq.~(\ref{Eq_dS}) in Appendix~\ref{Sec_AppendixST}). Firstly, we note that $\Delta S$ in our calculation is an increasing function of $M_\eta$ within the range $M_\eta\lesssim 400\,{\rm GeV}$, and secondly, $\Delta S\neq 0$ when $\sin\beta\rightarrow 0$.

\section{Applications}
\subsection{Mixing between $\eta:(7,4)$ and $\chi:(2,1)$}
We consider the model with an extra doublet $\chi=(\chi^+\;,\;{1\over\sqrt{2}}(\chi_R+i\chi_I))^T$ besides the septet $\eta$, where both scalars take odd charges under an $Z_2$ symmetry. The scalar potential can be written as
\begin{eqnarray}
-L&=&-\mu^2(\Phi^*\Phi)+\lambda (\Phi^*\Phi)^2+M_\chi^2 (\chi^*\chi)+\lambda_\chi (\chi^*\chi)^2\;,\nonumber\\
&&+{\lambda_5\over2}\Big[(\Phi^*\chi)^2+{\rm h.c.}\Big]+M_\eta^2(\eta^*\eta)
+\Big[{C_a\over \Lambda^3}\eta \chi(\Phi^*)^5+{C_b\over \Lambda^3}\eta \Phi(\Phi^*)^4\chi^*+{\rm H.c.}\Big]\;.\label{model_A}
\end{eqnarray}
Note that the $Z_2$ symmetry forbids the mixing of $\chi$ or $\eta$ with the SM doublet $\Phi$. Here, the quarter terms associated with $(\Phi^*\Phi)\chi^*\chi$ has been ignored, and the dimensional seven operators, $\eta \chi(\Phi^*)^5$ and $\eta \Phi(\Phi^*)^4\chi^*$, are imposed to yield the mixing between $\chi$ and $\eta$. Without the last term in Eq.~(\ref{model_A}), the model will be reduced to the ordinary inert doublet one~\cite{Deshpande:1977rw, Barbieri:2006dq, Ma:2006km, Majumdar:2006nt, LopezHonorez:2006gr, Hambye:2007vf, Andreas:2008xy}, in which the mass spectra are given by
\begin{eqnarray}
M_{\pm}^2=M_\chi^2\;,\;M_{R}^2=M_\chi^2+{\lambda_5\over 2}v^2\;,\; M_{I}^2=M_\chi^2-{\lambda_5\over 2}v^2\;.\;
\end{eqnarray}
The above masses turn into degeneracy when $\lambda_5$ approaches zero.
The mass splittings inside $\chi$ are the crucial quantities to give sizable contributions to $T$ or $S$. 
The deviation of the $T$ parameter for the inert doublet is~\cite{Barbieri:2006dq}
\begin{eqnarray}
\Delta T={1\over 16\pi m_W^2 s_W^2}[F(M_{\pm}^2,M_R^2)+F_(M_\pm^2,M_I^2)-F(M_R^2,M_I^2)]\;.
\end{eqnarray}
 where the definition of $F(x,y)$ is given in Appendix \ref{Sec_AppendixST}.
It is interesting to note that $\Delta T$ can be negative when there exists a large mass splitting between $\chi_R$ and $\chi_I$, which is governed by $M_\chi$ and $\lambda_5$. Note that in this case $\eta$ only contributes to $\Delta S$,  but not to $\Delta T$. 

The next step is to take into account the interaction between $\chi$ and $\eta$. We separately study the effects on masses of neutral and singly charged states with some non-zero values of $C_a$ and $C_b$, given by
\begin{eqnarray}
M_{x1}^2&=&{\cos^2 \theta_x\over \cos (2\theta_x)}M_x^2-{\sin^2 \theta_x\over \cos (2\theta_x)}M_\eta^2\;,\;\nonumber\\
M_{x2}^2&=&{\cos^2 \theta_x\over \cos (2\theta_x)}M_\eta^2-{\sin^2 \theta_x\over \cos (2\theta_x)}M_x^2\;,
\end{eqnarray}
where $M_{x1}$ and $M_{x2}$ are two mass eigenvalues of $\chi$ and $\eta$, and $\theta_x$ are mixing angles with $x=R(I)$ and $\pm$ corresponding to the real(imaginary) part of the neutral and charged components, respectively. When taking the condition $M_x<M_\eta$, it is obvious that $M_{x1}<M_{x2}$, which will be applied thereinafter. We list the mixing angles for two special cases as follows:
\begin{eqnarray}
(i)&:&C_a\neq0 \mbox{ and } C_b=0:\nonumber\\
&&\tan (2\theta_R)={{1\over\sqrt{6}}d_a\over M_R^2-M_\eta^2}v^2\;,\;
\tan (2\theta_I)={-{1\over\sqrt{6}}d_a\over M_I^2-M_\eta^2}v^2\;,\;
\tan (2\theta_\pm)={-d_a\over M_\pm^2-M_\eta^2}v^2\;;\nonumber\\
(ii)&:&C_a=0 \mbox{ and } C_b\neq0:\nonumber\\
&&\tan (2\theta_R)={{1\over\sqrt{6}}d_b\over M_R^2-M_\eta^2}v^2\;,\;
\tan (2\theta_I)={{1\over\sqrt{6}}d_b\over M_I^2-M_\eta^2}v^2\;,\;
\tan (2\theta_\pm)={{1\over\sqrt{15}}d_b\over M_\pm^2-M_\eta^2}v^2\;,\nonumber\\
\label{Eq_InnerMixing}
\end{eqnarray}
where $d_{j}=(C_{j}v^3)/(\sqrt{8}\Lambda^3)$ for $j=a,b$. The relevant coefficients in the above formulae can be found in Appendix \ref{Sec_AppendixMixing}. 
 Increasing the values of $d_{a,b}$ will also enlarge all the magnitudes of mixing angles between $\eta$ and $\chi$. For the limit of $\lambda_5\rightarrow 0$, we find that $|\theta_R|=|\theta_I|<|\theta_\pm|$ for (i), which means that $M_{R1}=M_{I1}>M_{\pm 1}$, so that the lightest inert component is $M_{\pm 1}$. Therefore, the stable neutral particle does not exist in this case.  On the other hand, the existence of the stable charged scalar makes this scenario diafovored by experiments, unless we further impose some other effective operator to break the $Z_2$ symmetry. The situation for (ii) is opposite as $|\theta_R|=|\theta_I|>|\theta_\pm|$ and $M_{R1}=M_{I1}<M_{\pm 1}$, resulting in a stable DM. Notice that in the above discussion the mass splitting scale generated by the mixing is usually larger than the quantum corrections by gauge bosons (see Ref.~\cite{Cirelli:2009uv}).

\begin{figure}
\includegraphics[width=16cm]{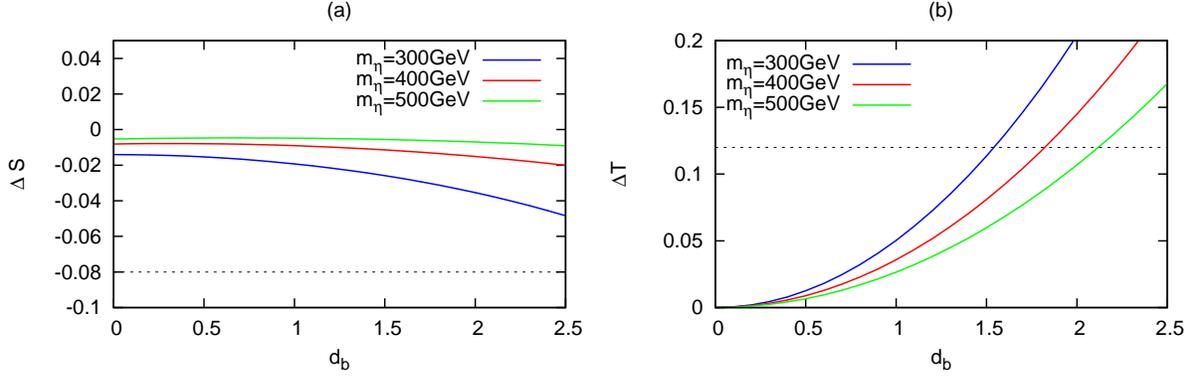} 
\caption{(a) $\Delta S$ and (b) $\Delta T$ as functions of $d_b$ with $M_\chi=250\,{\rm GeV}$ and different sets of $m_\eta$, where the horizontal dashed lines in (a) and (b) correspond to the $1.5-1.7\sigma$ bounds of $\Delta S\geq-0.08$ and $\Delta T\leq 0.12$~\cite{Agashe:2014kda}, respectively.}
\label{Fig_caseB}
\end{figure}

The explicit formulae for $\Delta T$ and $\Delta S$ are given in Appendix~\ref{Sec_AppendixST}. By fixing $M_\chi =250\,{\rm GeV}$, we plot them as functions of the coupling constant $d_b$ for (ii) in Fig.~\ref{Fig_caseB}. From the figure, we see that $\Delta S$ is always negative, whereas $\Delta T$ is positive. The magnitudes of $\Delta T$ raises with increasing $d_b$, and a larger $M_\eta$ gives smaller $\Delta S$ and $\Delta T$. For around $1.5\sigma$ deviation of an observed $\Delta T$, that is, $\Delta T<0.12$~\cite{Agashe:2014kda},  we have the upper bounds of $d_b<1.5,\;1.8$, and $2.1$ for $M_\eta=300,\,400,$ and $500\,{\rm GeV}$, respectively. On the other hands, the constraint by $\Delta S$ is relatively weak.

\subsection{Mixing between $\eta:(7,4)$ and $\rho:(1,4)$}
We now study the mixing of $\eta$ with a doubly-charge $SU(2)_L$ singlet $\rho:(1,4)$, which can have the lepton number $-2$ due to the coupling with the charged lepton $\rho\bar l_R^c l_R$. If $\eta^0$ acquires a VEV and $\rho$ mixes with $\eta^{++}$, then the lepton number is broken. Majorana neutrino masses can be generated through a two-loop diagram involving the interaction $\rho^{--}W^+W^+$~\cite{Chen:2006vn,Chen:2007dc,delAguila:2011gr,Chen:2012vm,King:2014uha}. To mix $\rho$ and $\eta$, like the scenario of the mixing between $\eta$ and $\Phi$, 
 some other higher dimensional operator are needed to achieve this goal in addition to the original one, $\eta\Phi(\Phi^5)^*$, which gives a nonzero $v_\eta$. A dimension-8 operator $\rho\eta \Phi^3(\Phi^3)^*$ is one of the possibility to drive the $\rho-\eta$ mixing. In this case, the potential is given by
\begin{eqnarray}
-L&=&-\mu^2(\Phi^*\Phi)+\lambda (\Phi^*\Phi)^2+M_\rho^2 (\rho^*\rho)\;,\nonumber\\
&&+M_\eta^2(\eta^*\eta)
+\Big[{1\over \Lambda^3}\eta \Phi(\Phi^*)^5+{1\over \Lambda'^4}\rho\eta^* (\Phi)^3(\Phi^*)^3+{\rm H.c.}\Big]\;,\label{Eq_doublyLagrangian}
\end{eqnarray}
where $\Lambda'$ is the mass scale related to the dimension-8 operator. Here, some of the quartic terms have been  ignored. We will focus on the discussion of the effects on the oblique parameters by the mixing pattern independent of the source of the mixture.

The doubly-charge mixing can be parametrized as:
\begin{eqnarray}
\left(
\begin{array}{c}
\rho^{++}\\
\eta^{++}\\
\end{array}\right)
=
\left(
\begin{array}{cc}
\cos\theta&-\sin\theta\\
\sin\theta&\cos\theta\\
\end{array}\right)
\left(
\begin{array}{c}
P_1^{++}\\
P_2^{++}\\
\end{array}\right)\;,\;
\label{Eq_doublyMixing}
\end{eqnarray}
where $\theta$ is the mixing angle and $P_{1,2}$ are the two mass eigenstates. The mass eigenvalues are then derived directly by
\begin{eqnarray}
M_{P_1}^2&=&{\cos^2 \theta\over \cos (2\theta)}M_\rho^2-{\sin^2 \theta\over \cos (2\theta)}M_\eta^2\;,\;\nonumber\\
M_{P_2}^2&=&{\cos^2 \theta\over \cos (2\theta)}M_\eta^2-{\sin^2 \theta\over \cos (2\theta)}M_\rho^2\;.
\end{eqnarray}
To study the influence on the electroweak structure, it is important to distinguish the deviations of the oblique parameters from different sources. Hence, The deviation of $T$ can be decomposed into two parts, $\Delta T=\Delta T_1+\Delta T_2$, where $\Delta T_1$ corresponds to the contribution from the $\eta-\Phi$ mixing, which is the same as the result in Eq.~(\ref{Eq_dT}), while $\Delta T_2$ is the rest given from the doubly-charge mixing, given by

\begin{eqnarray}
\Delta T_2&=&{1\over 4\pi s_W^2}\Big\{
15s_\beta^2[s_\theta^2 K(M_{P_1}^2,M_W^2)+c_\theta^2 K(M_{P_2}^2,M_W^2)-K(M_{\eta}^2,M_W^2)]\nonumber\\
&&+{6\over m_W^2}\Big[s_\theta^2F(M_{P_1}^2,M_\eta^2) + c_\theta^2F(M_{P_2}^2,M_\eta^2)\nonumber\\
&&+{5\over8}s_\beta^2(s_\theta^2F(M_{P_1}^2,M_{W}^2) + c_\theta^2F(M_{P_2}^2,M_{W}^2)
-F(M_\eta^2,M_{W}^2))\\
&&+{3\over 8}(s_\theta^2F(M_{P_1}^2,M_{\bar\eta^\pm}^2) + c_\theta^2F(M_{P_2}^2,M_{\bar\eta^\pm}^2)
)\nonumber\\
&&+{5\over8}c_\beta^2(s_\theta^2F(M_{P_1}^2,M_{S^\pm}^2) + c_\theta^2F(M_{P_2}^2,M_{S^\pm}^2)
-F(M_\eta^2,M_{S^\pm}^2))
\Big]
\Big\}\;.
\end{eqnarray}
The identical quantum numbers of $I_3$ and $Y$ between $\eta^{++}$ and $\rho^{++}$ make the relevant contribution to 
$\Delta T$ to be large because of the absence of $F(M_{P_1}^2,M_{P_2}^2)$. As a result, the mass splitting of doubly-charge 
eigenstates is constrained stringently. 
Similarly, $\Delta S=\Delta S_1+\Delta S_2$, where $\Delta S_1$ is the same as Eq.~(\ref{Eq_dS}), while $\Delta S_2$ is given by
\begin{eqnarray} 
\Delta S_2&=&-{2\over\pi}(4s_W^4)\Big[\xi({M_{1}^2\over M_Z^2},{M_{1}^2\over M_Z^2})
+\xi({M_{2}^2\over M_Z^2},{M_2^2\over M_Z^2})
-\xi({M_\eta^2\over M_Z^2},{M_\eta^2\over M_Z^2})\Big]\;.
\end{eqnarray}
It is obvious that the contribution from $\Delta S_2$ is tiny, because $I_3$ is zero for both $\rho^{\pm\pm}$ and $\eta^{\pm\pm}$, which makes the corresponding result proportional to $s_W^4$.

\begin{figure}
\includegraphics[width=16cm]{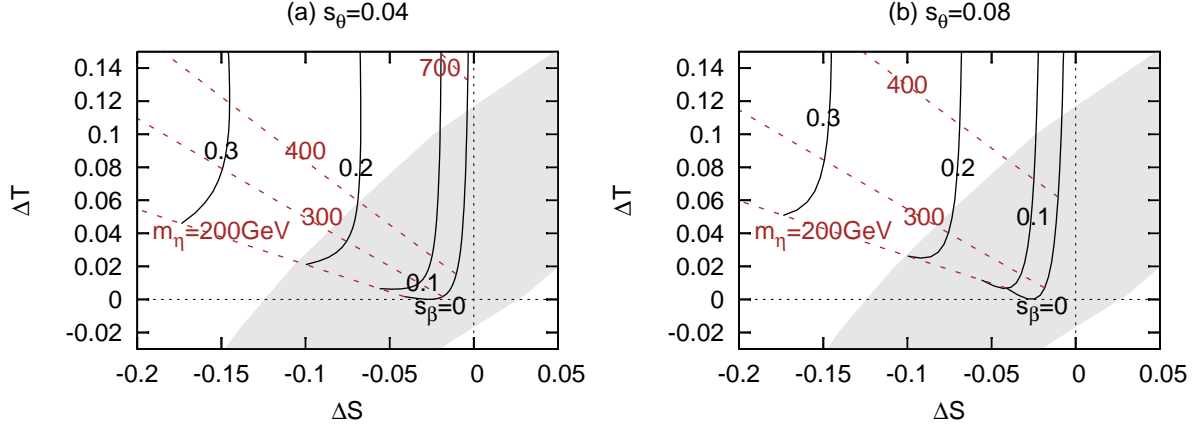}
\caption{Contours of $M_\eta$ and $\tan\beta$ in the $\Delta S-\Delta T$ planes, for (a) $\sin\theta=0.04$ and (b) $\sin\theta=0.08$ with $M_\rho={250{\rm GeV}}$, where the gray region corresponds to the global fitting results at $90\%$ C.L~\cite{Agashe:2014kda}.}
\label{Fig_doublySTplot}
\end{figure}

Our numerical results are shown in Fig.~\ref{Fig_doublySTplot}, where we have used $\sin\theta=0.04$ and $0.08$, together with $M_\rho=250\,{\rm GeV}$ as illustrations. When $M_\eta$ is large, say, $M_\eta\gtrsim 400\,{\rm GeV}$, $\Delta T_2$ can yield a significant contribution to $\Delta T$, so that the distortion in the $S-T$ plane appears, which is obvious in comparison with the ordinary figure in Fig.~\ref{Fig_STplot}. For a larger value of $\sin\theta$, the deformation is more apparent. Explicitly, we find that with $M_\eta=400\,{\rm GeV}$, $\sin\beta\lesssim 0.2\,(0.15)$ for $\sin\theta=0.04\,(0.08)$.

\section{Conclusions}
We have studied the model including a mixing term between the SM Higgs doublet and an extra septet $\eta:(7,4)$ under ${\rm SU}(2)_L\times {\rm U}(1)_Y$, which preserves the $\rho$ parameter to unity. The mixing between them arises from the effective dimension-7 operator $\eta \Phi(\Phi^*)^5$. The possible parameter space has been explored by examining the electroweak oblique parameters. We have focused on the mass range $M_\eta\lesssim 1500\,{\rm GeV}$. In general, $\Delta S$ is negative and increasing with $M_\eta$ under the range of $M_\eta\lesssim 400\,{\rm GeV}$, whereas $\Delta T$ is always positive and increasing for a large $\sin\beta$. 
Basically, the constraint on $\sin\beta$ changes slowly for different values of $M_\eta$. In the future, the global fitting of the oblique parameters could help to determine $M_\eta$ and $\tan\beta$ more precisely or directly exclude this model.

Besides the minimal doublet-septet mixing model, two extensions have been also discussed. One of them is to create an $Z_2$ odd sector which consists of the septet and an additional doublet $\chi$. Only two dimension-7 operators are possible to generate the mixing between $\eta$ and $\chi$. 
We have examined the effects of $\eta \Phi(\Phi^*)^4\chi^*$ on the scalar mass spectra and  oblique parameters. 
In particular, we have found the mixing coefficient $d_b\lesssim 1.8$ for $M_\eta=400\,{\rm GeV}$ and $M_\chi=250\,{\rm GeV}$. This result can help us to distinguish different mixing patterns by future analyses.
The other extension is to have an additional admixture between the septet and an SU(2) singlet  $\rho:(1,4)$. 
We can limit the mixing angle $\sin\theta$ in this kind of the models. For example, we have found that $\sin\beta\lesssim0.15$ for $M_\rho=250\,{\rm GeV}$, $\sin\theta=0.08$, and $M_\eta=400\,{\rm GeV}$, where the dominant constraint comes from $\Delta T$. The existence of this kind of the mixing pattern could also be tested by the oncoming experiments. 

\begin{acknowledgments}
This work was supported by National Center for Theoretical Sciences, National Science
Council (Grant No. NSC-101-2112-M-007-006-MY3) and National Tsing Hua University (Grant No. 104N2724E1).
\end{acknowledgments}

\appendix
\section{Mixing matrices for doublet-septet mixing model}\label{Sec_AppendixMixing}
To calculate the mass spectra in the doublet-septet mixing model, 
the first step is to deal with the non-hermitian effective coupling, which can be expanded as

\begin{eqnarray}
\eta \Phi(\Phi^*)^5+{\rm H.c.}&=&\eta^{(+5)} \Phi^0(\Phi^{*+})^5
+{1\over \sqrt{6}}\eta^{(+4)}[-({\Phi^+}^*)^5\Phi^++5(\Phi^{+*})^4\Phi^{0}]\nonumber\\
&&+{1\over \sqrt{15}}\eta^{+3}[-5({\Phi^+}^*)^4({\Phi^0}^*)\Phi^+ +10({\Phi^+}^*)^3({\Phi^0}^*)^2\Phi^0]\nonumber\\
&&+{1\over \sqrt{20}}\eta^{++}[-10({\Phi^+}^*)^3({\Phi^0}^*)^2\Phi^+ +10({\Phi^+}^*)^2({\Phi^0}^*)^3\Phi^0]\nonumber\\
&&+{1\over \sqrt{15}}\eta^{+}[-10({\Phi^+}^*)^2({\Phi^0}^*)^3\Phi^+ +5({\Phi^+}^*)({\Phi^0}^*)^4\Phi^0]\nonumber\\
&&+{1\over \sqrt{6}}\eta^0[-5({\Phi^+}^*)({\Phi^0}^*)^4\Phi^+ +({\Phi^0}^*)^5\Phi^0]
-\eta^-(\Phi^{0*})^5\Phi^{+}+{\rm h.c.}\nonumber\\
&\sim&{5\over \sqrt{15}}\eta^+{\Phi^{+}}^*\Phi^0({\Phi^{0}}^*)^4+{1\over\sqrt{6}}\eta^0\Phi^0(\Phi^{0*})^5-\eta^{-}\Phi^+({\Phi^0}^*)^5\nonumber\\
&&-{5\over\sqrt{6}}\eta^0({\Phi^+}^*)({\Phi^0}^*)^4\Phi^++{\rm H.c.}\;,
\end{eqnarray}
where we have only shown the relevant terms to 
provide the quadratic mixings of $\eta$ and $\Phi$ after the spontaneous symmetry breaking. The tadpole conditions in the potential are
\begin{eqnarray}
&&-\mu^2v_\Phi+\lambda v_\Phi^3+{12\over\sqrt{6}(\sqrt{2})^7\Lambda^3}v_\eta v_\Phi^5=0\;,\nonumber\\
&&M_\eta^2v_\eta+{2\over\sqrt{6}(\sqrt{2})^7\Lambda^3}v_\Phi^6=0\;.
\end{eqnarray}
The mixing matrix of the imaginary part is
\begin{eqnarray}
M_\eta^2\left(\begin{array}{cc}
{16v_\eta^2\over v_\Phi^2}&-{4v_\eta\over v_\Phi}\\
-{4v_\eta\over v_\Phi}&0\\
\end{array}\right)\;.
\end{eqnarray}
By comparing it with Eq.~(\ref{Eq_Neutralmixing}), we obtain the relation $t_\beta={4v_\eta/v_\Phi}$ and the eigenvalues $0$ and $M_\eta^2/c_\beta^2$. Similarly, the mass matrix of the neutral part is
\begin{eqnarray}
\left(\begin{array}{cc}
2\lambda c_\beta^2v^2-{3\over2}M_\eta^2t_\beta^2\;\;&-{3\over2}t_\beta M_\eta^2\\
-{3\over2}t_\beta M_\eta^2&M_\eta^2\\
\end{array}\right)\;.\label{Eq_realpartmatrix}
\end{eqnarray}
From Eqs.~(\ref{Eq_Neutralmixing}) and ~(\ref{Eq_realpartmatrix}), we get two relations:
\begin{eqnarray}
M_h^2s_\alpha^2+M_H^2c_\alpha^2=M_\eta^2\;,\;
c_\alpha s_\alpha(M_h^2-M_H^2)=-{3\over 2}M_\eta^2t_\beta\;,\;
\end{eqnarray}
which lead to the expressions of $M_h^2$ and $M_H^2$ shown in Eq.~(\ref{Eq_MassEigenvalues}).

The singly-charge mass matrix in the weak basis $\{\Phi^+\;,\;\eta^+\;,\;(\eta^-)^*\}$ is given by
\begin{eqnarray}
M_\eta^2\left(\begin{array}{ccc}
t_\beta^2&-{\sqrt{10}\over4}t_\beta&{\sqrt{6}\over4}t_\beta\\
-{\sqrt{10}\over4}t_\beta&1&0\\
{\sqrt{6}\over4}t_\beta&0&1\\
\end{array}\right)\;.\label{Eq_chargedmatrix}
\end{eqnarray}
The three eigenvalues of the matrix in Eq.~(\ref{Eq_chargedmatrix}) are $M_{G^\pm}^2=0$, $M_{\bar\eta^\pm}^2=M_\eta^2$, and $M_{S^\pm}^2=M_\eta^2/c_\beta^2$ with the transformation matrix
\begin{eqnarray}
U=\left(\begin{array}{ccc}
c_\beta&0&s_\beta\\
{\sqrt{10}\over4}s_\beta&{\sqrt{6}\over4}&-{\sqrt{10}\over4}c_\beta\\
-{\sqrt{6}\over4}s_\beta&{\sqrt{10}\over4}&{\sqrt{6}\over4}c_\beta\\
\end{array}\right)\;.
\end{eqnarray}
We now show the mixing pattern between the inert doublet and septet  discussed in Sec.~3. 
We have two types of the mixing models: (i) $\eta \chi(\Phi^*)^5$, and (ii) $\eta \Phi(\Phi^*)^4\chi^*$. 
Like the procedure in the beginning of this section, we extract the terms relevant to the $\chi-\eta$ mixing for (i) and (ii):
\begin{eqnarray}
&&(i):\;\;\eta \chi(\Phi^*)^5+{\rm h.c.}
\sim{1\over\sqrt{6}}\eta^0(\chi^0){(\Phi^0)^*}^5-\eta^-(\chi^+){(\Phi^0)^*}^5+{\rm H.c.}\;,\\
&&(ii):\;\;\eta \Phi(\Phi^*)^4\chi^*+{\rm h.c.}
\sim{1\over\sqrt{6}}\eta^0(\chi^0)^*\Phi^0{(\Phi^0)^*}^4+{1\over\sqrt{15}}\eta^-(\chi^+)^*\Phi^0{(\Phi^0)^*}^4+{\rm H.c.}\;.
\end{eqnarray}
Then, the relations in Eq.~(\ref{Eq_InnerMixing}) can be obtained directly.

\section{oblique parameters $\Delta S$ and $\Delta T$}\label{Sec_AppendixST}
In the doublet-septet mixing model, $\Delta S$ is given by
\begin{eqnarray}\label{Eq_dS}
 \Delta S&=&-{2\over\pi}\Big[\Big((3c_W^2-2s_W^2)^2+(2c_W^2-2s_W^2)^2+(c_W^2-2s_W^2)^2
 +4s_W^4\Big)\xi\Big({M_\eta^2\over M_Z^2},{M_\eta^2\over M_Z^2}\Big)\nonumber\\
&&+\Big({3\over2}-s_W^2\Big)^2\xi\Big({M_{\bar\eta^\pm}^2\over M_Z^2},{M_{\bar\eta^\pm}^2\over M_Z^2}\Big)
+\Big({1\over2}-s_W^2\Big)^2\xi\Big({M_{S^\pm}^2\over M_Z^2},{M_{S^\pm}^2\over M_Z^2}\Big)\Big]\nonumber\\
&&-{2\over\pi}\Big[{15\over2}s_\beta^2\xi\Big({M_W^2\over M_Z^2},{M_{\bar\eta^\pm}^2\over M_Z^2}\Big)
+{15\over2}c_\beta^2\xi\Big({M_{\bar\eta^\pm}^2\over M_Z^2},{M_{S^\pm}^2\over M_Z^2}\Big)\nonumber\\
&&+{1\over4}\Big(((c_\alpha c_\beta+4s_\alpha s_\beta)^2-1)\xi\Big({M_h^2\over M_Z^2},1\Big)
+(c_\alpha s_\beta-4s_\alpha c_\beta)^2\xi\Big({M_h^2\over M_Z^2},{M_{A}^2\over M_Z^2}\Big)\nonumber\\
&&+(s_\alpha c_\beta-4c_\alpha s_\beta)^2\xi\Big({M_H^2\over M_Z^2},1\Big)
+(s_\alpha s_\beta+4c_\alpha c_\beta)^2\xi\Big({M_H^2\over M_Z^2},{M_{A}^2\over M_Z^2}\Big)\Big)
\Big]\nonumber\\
&&-{1\over 3\pi}\Big[12 \log{M_\eta^2}
-{3\over4}\log{M_{\bar\eta^\pm}^2}+{1\over4}\log{M_{S^\pm}^2}
-{15\over4}s_\beta^2\log({M_{W}^2M_{\bar\eta^\pm}^2})-{15\over4}c_\beta^2\log(M_{\bar\eta^\pm}^2M_{S^\pm}^2)
\nonumber\\
&&-{1\over8}((c_\alpha c_\beta+4s_\alpha s_\beta)^2-1)\log(M_{h}^2M_{Z}^2)
-{1\over8}(c_\alpha s_\beta-4s_\alpha c_\beta)^2\log(M_{h}^2M_{A}^2)\nonumber\\
&&-{1\over8}(s_\alpha c_\beta-4c_\alpha s_\beta)^2\log(M_{H}^2M_{Z}^2)
-{1\over8}(s_\alpha s_\beta+4c_\alpha c_\beta)^2\log(M_{H}^2M_{A}^2)\Big]\nonumber\\
&&+{1\over \pi}\Big[30s_\beta^2c_W^2\zeta\Big({M_{\bar\eta^\pm}^2\over M_Z^2},{M_W^2\over M_Z^2}\Big)
+[(c_\alpha c_\beta +4s_\alpha s_\beta)^2-1]\zeta\Big({M_h^2\over M_Z^2},1\Big)\nonumber\\
&&+(-s_\alpha c_\beta+4 c_\alpha s_\beta)^2\zeta\Big({M_H^2\over M_Z^2},1\Big)
\Big]\;, 
\end{eqnarray}
where~\cite{Lavoura:1993nq}
\begin{eqnarray}
\xi(x,y)&=&{4\over9}-{5\over12}(x+y)+{1\over6}(x-y)^2+{1\over4}\Big[x^2-y^2-{1\over3}(x-y)^3-{x^2+y^2\over x-y}\Big]\log{x\over y}\nonumber\\
&&-{1\over12}\Delta(x,y)f(x,y)\;,\\
\zeta(x,y)&=&{1\over2}\Big[{x-y}-{x+y\over x-y}\Big]\log{x\over y}-1-{1\over2}f(x,y)\;,\\
\Delta(x,y)&=&-1+2(x+y)-(x-y)^2\;,
\end{eqnarray}
\begin{eqnarray}
f(x,y)=\left\{\begin{array}{lc}
-2\sqrt{\Delta(x,y)}\Big(\tan^{-1}{x-y+1\over\sqrt{\Delta(x,y)}}-\tan^{-1}{x-y-1\over\sqrt{\Delta(x,y)}}\Big)&\mbox{ for }\Delta(x,y)>0\;,\\
0&\mbox{ for }\Delta(x,y)=0\;,\\
\sqrt{-\Delta(x,y)}\log{x+y-1+\sqrt{-\Delta(x,y)}\over x+y-1-\sqrt{-\Delta(x,y)}}&\mbox{ for }\Delta(x,y)<0\;.
\end{array}\right.
\end{eqnarray}
Note that the above formula for $\Delta S$ is different from that given in Ref.~\cite{Hisano:2013sn}.  
In comparison with the formulae in Ref.~\cite{Hisano:2013sn}, we have one additional contribution from the first term in Eq.~(\ref{Eq_dS}), which decreases with a larger value of $M_\eta$, but it is dominated when $\sin\beta$ is small. However, the result in Ref.~\cite{Hisano:2013sn} could be compatible with ours when $M_\eta$ is not too small. It is easily checked that when $\sin\beta\rightarrow 0$, our result of $\Delta S$ from Eq.~(\ref{Eq_dS}) is nonzero, whereas $\Delta S \rightarrow 0$ in Ref.~\cite{Hisano:2013sn}.
 $\Delta T$ is found to be~\cite{Hisano:2013sn}
\begin{eqnarray}\label{Eq_dT}
\Delta T&=&
{1\over 4\pi s_W^2M_W^2}\Big\{{1\over4}((c_\alpha c_\beta +4s_\alpha s_\beta)^2-1)(G(M_{h}^2,M_{W}^2)-G(M_{h}^2,M_{Z}^2))\nonumber\\
&&+{1\over4}(-s_\alpha c_\beta +4c_\alpha s_\beta)^2(G(M_{H}^2,M_{W}^2)-G(M_{H}^2,M_{Z}^2))\nonumber\\
&&-{15s_\beta^2\over4}(G(M_{\eta}^2,M_{W}^2)-G(M_{\eta}^2,M_{Z}^2))\Big]\;,\nonumber\\
\end{eqnarray}
where
\begin{eqnarray}
G(x,y)&=&F(x,y)+4yK(x,y)\;,\;\\
F(x,y)&=&{x+y\over2}-{xy\over x-y}\log{x\over y}\;,\;
K(x,y)={x\log x-y\log y\over x-y}\;.\;
\end{eqnarray}
In this model, $\Delta U$ is given by
\begin{eqnarray}
\Delta U&=&{2\over \pi}\Big[ \Big( (3c_W^2-2s_W^2)^2+(2c_W^2-2s_W^2)^2+(c_W^2-2s_W^2)^2+(-2s_W^2)^2 \Big)
\xi\Big({M_\eta^2\over M_Z^2},{M_\eta^2\over M_Z^2}\Big)
-{7\over 6}\log(M_\eta^4)\nonumber\\
    &&+ ({3\over2}-s_W^2)^2\log\Big({M_{\bar\eta}^2\over M_Z^2},{M_{\bar\eta}^2\over M_Z^2}\Big)
    -{3\over 16}\log(M_{\bar\eta}^4)
    + ({1\over2}-s_W^2)^2\xi\Big({M_S^2\over M_Z^2},{M_S^2\over M_Z^2}\Big)
    -{1\over 48}\log(M_S^4)\nonumber\\
    &&+{15\over2}s_\beta^2\Big(\xi\Big({M_W^2\over M_Z^2},{M_{\bar\eta}^2\over M_Z^2}\Big)-2c_W^2\zeta\Big({M_W^2\over M_Z^2},{M_{\bar \eta}^2\over M_Z^2}\Big)-{1\over 12}\log(M_W^2 M_{\bar\eta}^2)\Big) \nonumber\\     
&&+{15\over2}c_\beta^2\Big(\xi\Big({M_{\bar\eta}^2\over M_Z^2},{M_S^2\over M_Z^2}\Big)-{1\over 12}\log(M_S^2M_{\bar\eta}^2)\Big)
\Big]\nonumber\\
&&+{1\over 2\pi}\Big[ 
       ((c_\alpha c_\beta+4s_\alpha s_\beta)^2-1)\Big(\xi\Big({m_h^2\over M_Z^2},1\Big)-2\zeta\Big({m_h^2\over M_Z^2},1\Big)-{1\over 12}\log(M_h^2M_Z^2)\Big)\nonumber\\
       &&+(-s_\alpha c_\beta+4c_\alpha s_\beta)^2\Big(\xi\Big({M_H^2\over M_Z^2},1\Big)-2\zeta\Big({M_H^2\over M_Z^2},1\Big)-{1\over 12}\log(M_H^2M_Z^2)\Big)\nonumber\\
       &&+(-c_\alpha s_\beta+4s_\alpha c_\beta)^2\Big(\xi\Big({M_h^2\over M_Z^2},{M_A^2\over M_Z^2}\Big)-{1\over 12}\log(M_h^2M_A^2)\Big)\nonumber\\
       &&+(s_\alpha s_\beta+4c_\alpha c_\beta)^2\Big(\xi\Big({M_H^2\over M_Z^2},{M_A^2\over M_Z^2}\Big)-{1\over 12}\log(M_H^2M_A^2)\Big)\Big]\nonumber\\
&&-{28\over \pi}
\Big(\xi\Big({M_\eta^2\over M_W^2},{M_\eta^2\over M_W^2}\Big)
-{1\over 12}\log(M_\eta^4)\Big)\nonumber\\
&&-{1\over2\pi}\Big[((c_\beta c_\alpha+4s_\beta s_\alpha)^2-1)\Big(\xi\Big(1,{M_h^2\over M_W^2}\Big)-{1\over12}\log(M_W^2 M_h^2)-2\zeta\Big(1,{M_h^2\over M_W^2}\Big) \Big)\nonumber\\
&&+(-c_\beta s_\alpha+4s_\beta c_\alpha)^2\Big(\xi\Big(1,{M_H^2\over M_W^2}\Big)-{1\over12}\log(M_W^2 M_H^2)-2\zeta\Big(1,{M_H^2\over M_W^2}\Big) \Big)\nonumber\\
&&+15s_\beta^2\Big(\xi\Big({M_{\bar\eta}^2\over M_W^2},{M_Z^2\over M_W^2}\Big)-{1\over12}\log(M_{\bar\eta}^2 M_Z^2)-{2\over c_W^2}\zeta\Big({M_{\bar\eta}^2\over M_W^2},{M_Z^2\over M_W^2}\Big) \Big)\nonumber\\
&&+15c_\beta^2\Big(\xi\Big({M_{\bar\eta}^2\over M_W^2},{M_A^2\over M_W^2}\Big)-{1\over12}\log(M_{\bar \eta}^2 M_A^2) \Big)\nonumber\\
&&+(s_\beta c_\alpha-4c_\beta s_\alpha)^2\Big(\xi\Big({M_S^2\over M_W^2},{M_h^2\over M_W^2}\Big)-{1\over12}\log(M_S^2 M_h^2)\Big)\nonumber\\
&&+(-s_\beta s_\alpha-4c_\beta c_\alpha)^2\Big(\xi\Big({M_S^2\over M_W^2},{M_H^2\over M_W^2}\Big)-{1\over12}\log(M_S^2 M_H^2)\Big)\nonumber\\
&&+\Big(\xi\Big({M_S^2\over M_W^2},{M_A^2\over M_W^2}\Big)-{1\over12}\log(M_S^2 M_A^2) \Big)\Big]\nonumber\\
    &&-{1\over\pi}\Big[{15\over2}s_\beta^2\Big(\xi\Big({M_\eta^2\over M_W^2},1\Big)-2\zeta\Big({M_\eta^2\over M_W^2},1\Big)-{1\over 12}\log(M_W^2 M_{\eta}^2)\Big) \nonumber\\     
&&+{9\over2}\Big(\xi\Big({M_\eta^2\over M_W^2},{M_{\bar\eta}^2\over M_W^2}\Big)-{1\over 12}\log(M_\eta^2 M_{\bar\eta}^2)\Big)
+{15\over2}c_\beta^2\Big(\xi\Big({M_\eta^2\over M_W^2},{M_{S}^2\over M_W^2}\Big)-{1\over 12}\log(M_\eta^2 M_{S}^2)\Big)\Big]\,.       
\end{eqnarray}

We now list $\Delta S$ and $\Delta T$ in the $\chi-\eta$ mixing model, given by
\begin{eqnarray}
\Delta S&=&-{2\over\pi}\Big[\Big(23-58s_W^2+31s_W^4\Big)\xi({M_\eta^2\over M_Z^2},{M_\eta^2\over M_Z^2})
+\Big(({1\over2}c_\pm^2-s_\pm^2-s_W^2)^2\xi({M_{\pm1}^2\over M_Z^2},{M_{\pm1}^2\over M_Z^2})\nonumber\\
&&+{9\over2}c_\pm^2s_\pm^2\xi({M_{\pm1}^2\over M_Z^2},{M_{\pm2}^2\over M_Z^2})
+({1\over2}s_\pm^2-c_\pm^2-s_W^2)^2\xi({M_{\pm2}^2\over M_Z^2},{M_{\pm2}^2\over M_Z^2})\Big)\nonumber\\
&&
+{1\over4}\Big((c_R c_I+4s_R s_I)^2\xi({M_{R1}^2\over M_Z^2},{M_{I1}^2\over M_Z^2})
+(c_R s_I-4s_R c_I)^2\xi({M_{R1}^2\over M_Z^2},{M_{I2}^2\over M_Z^2})\nonumber\\
&&+(s_R c_I-4c_R s_I)^2\xi({M_{R2}^2\over M_Z^2},{M_{I1}^2\over M_Z^2})
+(s_R s_I+4c_R c_I)^2\xi({M_{R2}^2\over M_Z^2},{M_{I2}^2\over M_Z^2})\Big)
\Big]\nonumber\\
&&-{1\over 3\pi}\Big[6 \log{M_\eta^2}
+\Big(({1\over4}-{9\over4}s_\pm^4)\log{M_{\pm1}^2}
-{9\over4}c_\pm^2s_\pm^2\log{M_{\pm1}^2M_{\pm2}^2}
+({1\over4}-{9\over4}c_\pm^4)\log{M_{\pm2}^2}\Big)
\nonumber\\
&&-{1\over8}\Big((c_Rc_I+4s_Rs_I)^2\log{M_{R1}^2M_{I1}^2}
+(c_Rs_I-4s_Rc_I)^2\log{M_{R1}^2M_{I2}^2}\nonumber\\
&&+(s_Rc_I-4c_Rs_I)^2\log{M_{R2}^2M_{I1}^2}
+(s_Rs_I+4c_Rc_I)^2\log{M_{R2}^2M_{I2}^2}\Big)\Big]\;,\nonumber\\
\end{eqnarray}

\begin{eqnarray}
\Delta T&=&{1\over 4 \pi M_W^2 s_W^2 }\Big\{
6[s_{\theta_\pm}^2F(M_\eta^2,M_{\pm1}^2)+c_{\theta_\pm}^2F(M_\eta^2,M_{\pm2}^2)]\nonumber\\
&&+{1\over4}\sum_{X=R,I}\Big[(c_\pm c_X+\sqrt{10} s_\pm s_X)^2 F(M_{\pm1}^2,M_{X1}^2)+
(-c_\pm s_X+\sqrt{10} s_\pm c_X)^2 F(M_{\pm1}^2,M_{X2}^2)\nonumber\\
&&+(-s_\pm c_X+\sqrt{10} c_\pm s_X)^2 F(M_{\pm2}^2,M_{X1}^2)+
(s_\pm s_X+\sqrt{10} c_\pm c_X)^2 F(M_{\pm2}^2,M_{X_2}^2)\Big]\nonumber\\
&&+{3\over2}[s_R^2F(M_{R1}^2,M_{\eta}^2)+c_R^2F(M_{R2}^2,M_{\eta}^2)
+s_I^2F(M_{I1}^2,M_{\eta}^2)+c_I^2F(M_{I2}^2,M_{\eta}^2)]\nonumber\\
&&-{9\over2}c_\pm^2s_\pm^2F(M_{\pm1}^2,M_{\pm2}^2)\nonumber\\
&&-{1\over4}\Big[(c_Rc_I+4s_Rs_I)^2F(M_{R1}^2,M_{I1}^2)
+(-c_Rs_I+4s_Rc_I)^2F(M_{R1}^2,M_{I2}^2)\nonumber\\
&&+(-s_Rc_I+4c_Rs_I)^2F(M_{R2}^2,M_{I1}^2)
+(s_Rs_I+4c_Rc_I)^2F(M_{R2}^2,M_{I2}^2)\Big]
\Big\}\;.
\end{eqnarray}

\end{document}